\documentclass[12pt]{iopart}
\usepackage{iopams}           
\usepackage{graphicx}         
\usepackage{bm}               
\usepackage{lineno}           

\usepackage[usenames]{color}               
\definecolor{MyDarkRed}{rgb}{0.30,0.08,0.0} 
\definecolor{MyDarkGreen}{rgb}{0.0,0.30,0.08} 
\definecolor{MyDarkBlue}{rgb}{0.0,0.08,0.30} 

\usepackage[
    bookmarks=true,         
    unicode=true,          
    pdftoolbar=true,        
    pdfmenubar=true,        
    pdffitwindow=true,      
    pdftitle={A Three-Point Cosmic Ray Anisotropy Method},    
    pdfauthor={J. D. Hague},     
    pdfsubject={Astroparticle Physics},   
    pdfcreator={J. D. Hague},   
    pdfproducer={J. D. Hague}, 
    pdfkeywords={cosmic-ray anisotropy}, 
    colorlinks=true,        
    linkcolor=MyDarkBlue,          
    citecolor=MyDarkRed,        
    filecolor=MyDarkGreen,      
    urlcolor=MyDarkGreen         
]{hyperref}
\usepackage[all]{hypcap}    

\begin{document}

\newcommand{\secref}[1]{\sref{#1}}
\newcommand{\figref}[1]{\Fref{#1}}
\newcommand{\equref}[1]{\eref{#1}}
\newcommand{\tabref}[1]{Table \ref{#1}}
\newcommand{\degree}{\ensuremath{^{\circ}}}
\newcommand{\Nskies}{{10^{4}}}
\newcommand{\SP}{{\Sigma_{P}}}
\newcommand{\tal}{{\alpha}}
\newcommand{\talo}{{\alpha_{01}}}
\newcommand{\taloo}{{\alpha_{001}}}
\newcommand{\tbl}{{\beta}}
\newcommand{\tblo}{{\beta_{\talo}}}
\newcommand{\tbloo}{{\beta_{\taloo}}}
\newcommand{\ho}{{\mathsf{H}_{iso}}}
\newcommand{\hs}{{\mathsf{H}_{sig}}}
\newcommand{\nobs}{{n_{obs}}}
\newcommand{\nexp}{{n_{exp}}}
\newcommand{\lnPi}{{\ln P_{i}(\nobs | \nexp)}}
\newcommand{\lnPt}{{\ln P_{\theta}(\nobs | \nexp)}}
\newcommand{\lnPss}{{\ln P_{\gamma \zeta}(\nobs | \nexp)}}
\newcommand{\arXiv}[2]{\href{#1}{{\tt #2}}}

\title[A Three-Point Cosmic Ray Anisotropy Method]{A Three-Point Cosmic Ray Anisotropy Method}
\author{J.~D.~Hague, B.~R.~Becker, M.~S.~Gold, J.A.J.~Matthews}
\address{
  University of New Mexico, Department of Physics and Astronomy \\
  Albuquerque, New Mexico, USA
}
\ead{jhague@unm.edu}

\begin{abstract}
The two-point angular correlation function is a traditional method used to search for deviations from expectations of isotropy. 
In this paper we develop and explore a statistically descriptive three-point method with the intended application being the 
search for deviations from isotropy in the highest energy cosmic rays.  
We compare the sensitivity of a two-point method and a ``shape-strength'' method for a variety of Monte-Carlo simulated anisotropic signals.  
Studies are done with anisotropic source signals diluted by an isotropic background. 
Type I and II errors for rejecting the hypothesis of isotropic cosmic ray arrival directions are evaluated for four
different event sample sizes: 27, 40, 60 and 80 events, consistent with near term data expectations from the Pierre Auger Observatory. 
In all cases the ability to reject the isotropic hypothesis improves with event size and with the fraction of anisotropic signal. 
While $\sim 40$ event data sets should be sufficient for reliable identification of anisotropy in cases of rather extreme
(highly anisotropic) data, much larger data sets are suggested for reliable identification of more subtle anisotropies.
The shape-strength method consistently performs better than the two point method and can be easily adapted to an arbitrary experimental 
exposure on the celestial sphere. 
\end{abstract}

\vspace{2pc}
\noindent {\it Submitted to:} \JPG \\ 
\noindent {\it Dated:} \today
\maketitle

\section{Introduction} \label{sec:intro}
Cosmic rays with energies above 10 EeV ($10^{19}$ eV) have been observed\cite{Linsley:1963km,Takeda:1998ps,Abbasi:2007sv,Abraham:2008ru}.  
However, the sources of these cosmic rays (CR) are unknown and the physics responsible for accelerating 
CR to these energies is at best conjecture.  
Evidence supporting an extra-galactic origin of these CR is the observation 
of energy flux suppression consistent with the GZK-effect\cite{refG,refZK} by 
the High Resolution Fly's Eye Experiment\cite{Abbasi:2007sv} and 
the Pierre Auger Observatory (Auger)\cite{Abraham:2008ru}.
The primary evidence supporting the astrophysical origin of these CR (as opposed to, say, heavy relic decay) is the 
lack of an observable flux of photons by Auger\cite{Aglietta:2007yx,Abraham:2006ar,Collaboration:2009qb} and 
the lack of neutrinos observed by ANITA\cite{Barwick:2005hn,Gorham:2008yk}.   

If the sources are astrophysical, expectations for asymmetries in the arrival directions increase at the 
very highest CR energies because the local ($\lesssim100$ Mpc) universe is very anisotropic\cite{refVCV,refIRAS}
and the GZK-effect\cite{Abbasi:2007sv,Abraham:2008ru} at these energies implies that the sources are local.
Observation of an anisotropy in the arrival directions of CR would be an important step towards identifying the sources of 
these ultra high energy particles.

Evidence for structure (anisotropy) in the arrival directions has been reported
\cite{PhysRevLett.77.1000,Takeda:1999sg,Kachelriess:2005uf,Jansson:2007xi,Cronin:2007zz,Abraham:2007si,Mollerach:2007vb,Mollerach:2009tp}. 
The most compelling observational evidence consistent with astrophysical expectations of anisotropy is arguably the 27 events with energy greater than
57 EeV recently reported by Auger in \cite{Cronin:2007zz, Abraham:2007si}. 
Using the V\'{e}ron-Cetty -- V\'{e}ron (VCV) catalog\cite{refVCV}, the active galactic nuclei (AGN) maximum redshift and correlation angle chosen by Auger 
defined a limited area (effectively 21\%) of the sky\cite{Abraham:2007si}. 
Reported at the 1\% significance level, 
the Auger AGN to CR correlation signal is evidence for a flux of CR enhanced near known low-redshift extra-galactic objects\cite{Abraham:2007si}. 

As even the largest experiments accumulate the very highest energy CR only slowly,
\footnote{For example, the Auger event rate for CR above the GZK knee, $\sim 56$ EeV, 
  is on the order of two per month\cite{Abraham:2008ru}.} 
the development of new, more sensitive, techniques to search for deviations from isotropy is of particular interest. 
In contrast to the catalog dependent method used by \cite{Cronin:2007zz,Abraham:2007si,Koers:2008ba}, 
in this paper we study the effectiveness of two catalog independent methods. Catalog independent techniques avoid the 
penalty factors for scans over many different catalogs and/or the need to restrict the CR data
based on limited sky coverage of a catalog. 
The first catalog independent technique is a binned two-point ({\it 2-Pt}) angular correlation method (\secref{ssec:2pt}).
We also introduce a new three-point method which uses a shape and a strength parameter ({\it S-S}, or Shape-Strength) 
for each triplet of events (\secref{ssec:ss}). 
Both methods are compared throughout via the binned-likelihood analysis described in \secref{ssec:ap}.

Arguably, the primary impediment to definitive CR source identification is the small number of ultra-high energy events 
(those near and above the GZK cut-off, which are most likely to be anisotropic).
While lower energy events are more abundant, their sources are likely to be further away, where the universe is isotropic. 
Furthermore, galactic/intergalactic magnetic fields are likely to wash out any correlation with the sources of lower energy events 
(neglecting the possible effects of magnetic field caustics\cite{Harari:2002fj,Harari:2006xx}).
Thus, as one decreases the minimum observed energy one expects to include events which dilute any high energy anisotropy signal. 
Furthermore, there is typically significant error in the value of an observed energy (as much as $25\%$\cite{Abraham:2008ru}).
We therefore pay careful attention to the performance of the methods under variation in the total number of events and dilution 
factor (signal to isotropic background) for different types of signals in \secref{ssec:ms}. 

\section{Methods} \label{sec:meth}
The 2-Pt (\secref{ssec:2pt}) and S-S (\secref{ssec:ss}) methods are compared using the analysis paradigm described in \secref{ssec:ap} 
When needed for a concrete example, we use the largest currently operating observatory (Auger) for representative data set sizes 
and sky exposure\cite{Cronin:2007zz, Abraham:2007si}.
The methods presented here, however, can be applied to a spherical data set of any size and with an arbitrary experimental exposure.

\subsection{Analysis Paradigm} \label{ssec:ap}
We use a similar analysis paradigm for both the 2-Pt and S-S methods to calculate a $p$-value for 
rejecting the isotropic (null) hypothesis, $\ho$. 
Each method uses binned {\it parameters} to compute a pseudo-log-likelihood 
test statistic $\SP$, ``pseudo-'' because the bins are correlated.
The correlation does not effect the final answer because the $p$-value is 
derived by comparing the distribution of the $\SP$ in a test sky to that of identically analyzed isotropic skies. 
The flatness of the distribution of p-values for isotropic test skies has been verified. 
The parameter space for the 2-Pt method is the angular distance between two events. 
For the S-S method the parameter space is two dimensional. 
In neither case is the parameter space scanned to determine an optimal value. 
Instead, we compare the entire observed distribution to that expected by isotropy. 

For a given set of cosmic ray events (referred to here as a {\it sky}) we compute $\SP$ by comparing the 
binned distribution of the test sky's parameter(s) to the parameter(s) distribution expected from an isotropic sky.
The probability for observing $\nobs$ doublets (2-Pt) or triplets (S-S) from the test sky in the $i^{th}$ parameter bin, 
given that you expect\cite{isoexp} 
to see $\nexp$ from an isotropic sky, is approximated\cite{poistail} 
by a Poisson distribution $P_{i}(\nobs | \nexp) = \nexp^{\nobs}e^{-\nexp} / \nexp!$.
The pseudo-log-likelihood is $\SP = \sum_{i} \lnPi$, 
where the sum is carried out over the bins of the parameter space.
The ratio of the number of isotropic skies with $\SP$ less than that of the test sky to the total number of simulated 
isotropic skies gives the $p$-value for the test sky.

In the following discussion $\vec{r}_{k}$ is defined as the arrival direction of the $k^{th}$ event in a sky.  
This (unit) vector has Cartesian coordinates $\{r_{x}, r_{y}, r_{z}\}$ when projected from the galactic sphere. 

\subsection{Two-Point Correlation} \label{ssec:2pt}
The 2-Pt correlation distribution is calculated by computing the number 
of event pairs in a test sky as a function of the angular distance between any two events, 
$\theta = \cos^{-1}(\vec{r}_{j} \cdot \vec{r}_{k})$ (see \cite{Mollerach:2007vb,Mollerach:2009tp,Cuoco:2007id} for similar methods). 
We use $5\degree$ bins for $\theta \in [0\degree, 180\degree)$, so that the pseudo-log-likelihood is the sum over all 
angular scales, $\SP = \sum_{\theta} \lnPt$. 

\subsection{Shape-Strength} \label{ssec:ss}
This method involves an eigenvector decomposition, or principle component analysis, 
of the arrival directions of all sets of triplets found in the data set. 
It is inspired primarily by Fisher \cite{refFisher} (see also \cite{refWood1,refWood2}) 
but differs in that we decompose all subsets of triplets in a sky to obtain a test statistic. 

\begin{figure}[ht]
  \begin{tabular}{cc}
    \includegraphics[width=0.35\linewidth]{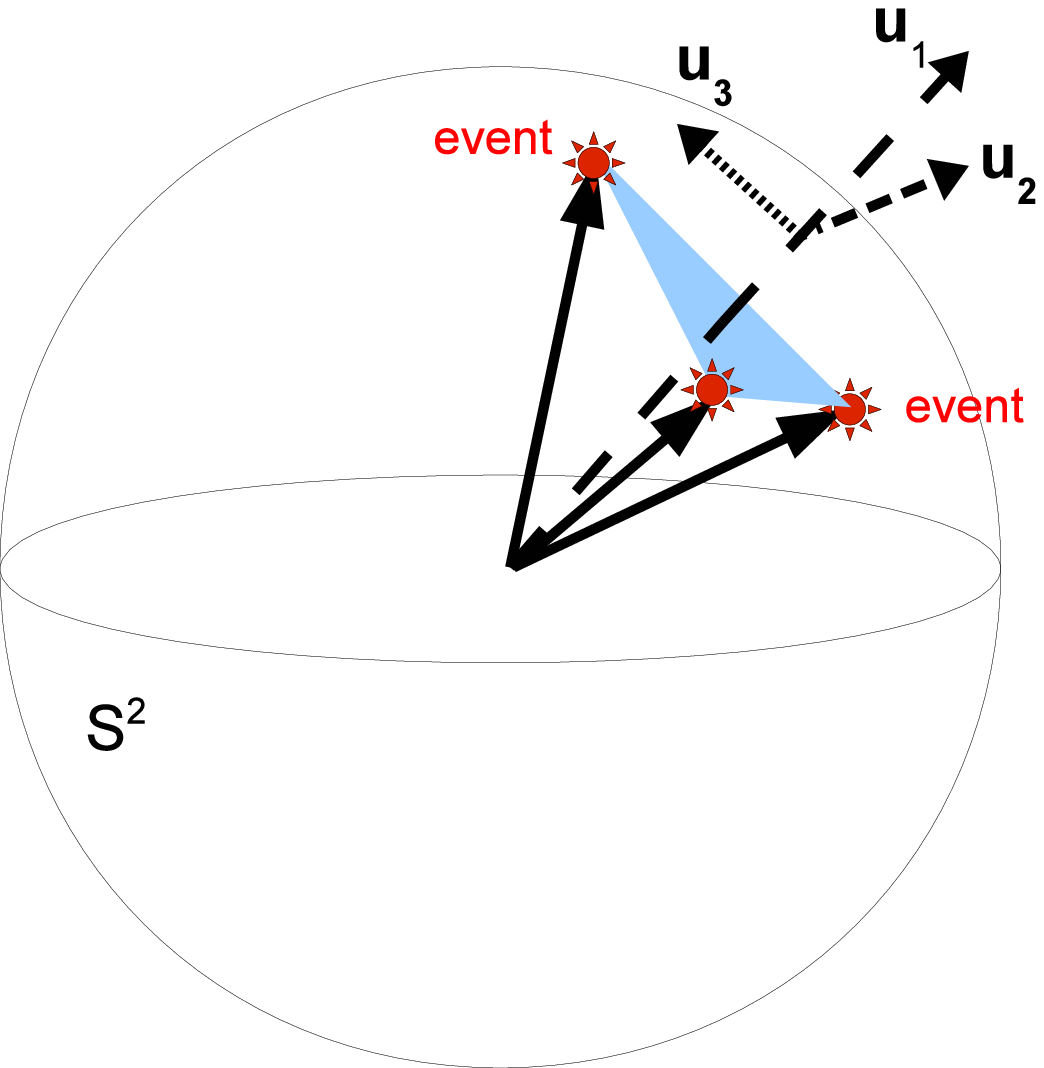} & 
    \includegraphics[width=0.65\linewidth]{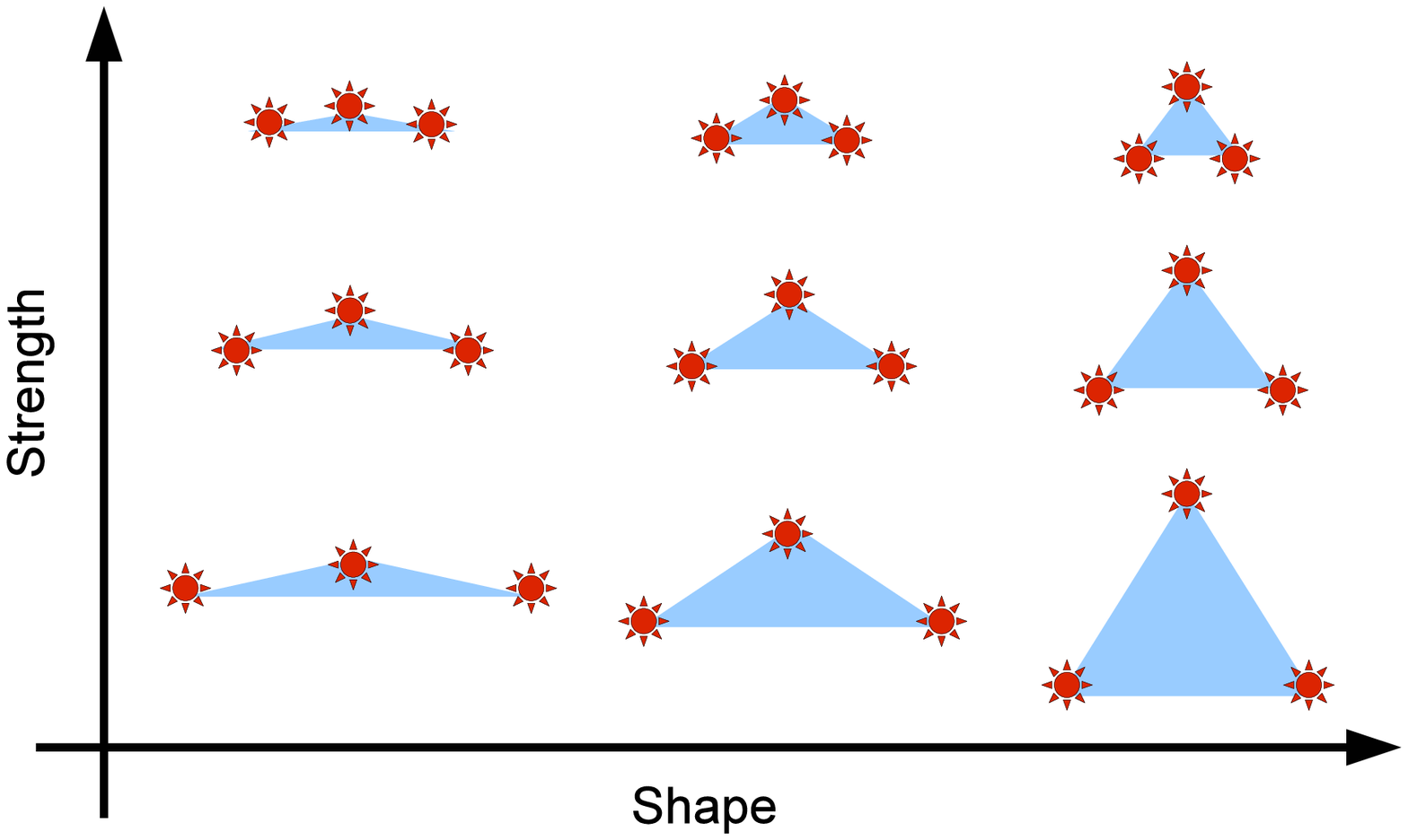} 
  \end{tabular}
   \caption{\label{fig:trip}
     {\it Left:} The eigenvectors of a triplet of events on the sphere ($S^{2}$) are the principle axis $\vec{u}_{1}$, 
     the  major axis $\vec{u}_{2}$ (pointing into the page) and the minor axis $\vec{u}_{3}$.
     The eigenvalues of these vectors are used to compute this triplet's shape and strength.
     {\it Right:} An intuitive interpretation of the shape and strength parameters.
     As the strength parameter $\zeta$ increases from $0$ to $\infty$, the events become more concentrated.
     As the shape parameter $\gamma$ increases form $-\infty$ to $+\infty$, the events become more rotationally symmetric or less elongated. 
}
\end{figure}

For each triplet we calculate the components of the symmetric ($3 \times 3$) orientation 
matrix $\mathbf{T}$\cite{refFisher}. 
In Cartesian coordinates, $T_{ij} = \frac{1}{3}  \sum_{k \in triplet} \left(  r_{i}r_{j} \right)_{k}$ for $i,j \in \{x,y,z\}$.
The largest eigenvalue of $\mathbf{T}$, $\tau_{1}$, results from a rotation of the triplet about the {\it principle} axis $\vec{u}_{1}$.
The middle and smallest eigenvalues correspond to the {\it major} $\vec{u}_{2}$ and {\it minor} $\vec{u}_{3}$ axis respectively. 
The left panel of \figref{fig:trip} shows a graphical illustration of these eigenvectors.
The eigenvalues satisfy $\tau_{1} + \tau_{2} + \tau_{3} = 1$ and $\tau_{1}\geq\tau_{2}\geq\tau_{3}\geq0$, 
and thus there are only two independently varying parameters for any triplet.

\begin{figure*}[htp]
  \begin{tabular}{cc}
    \includegraphics[width=0.50\linewidth]{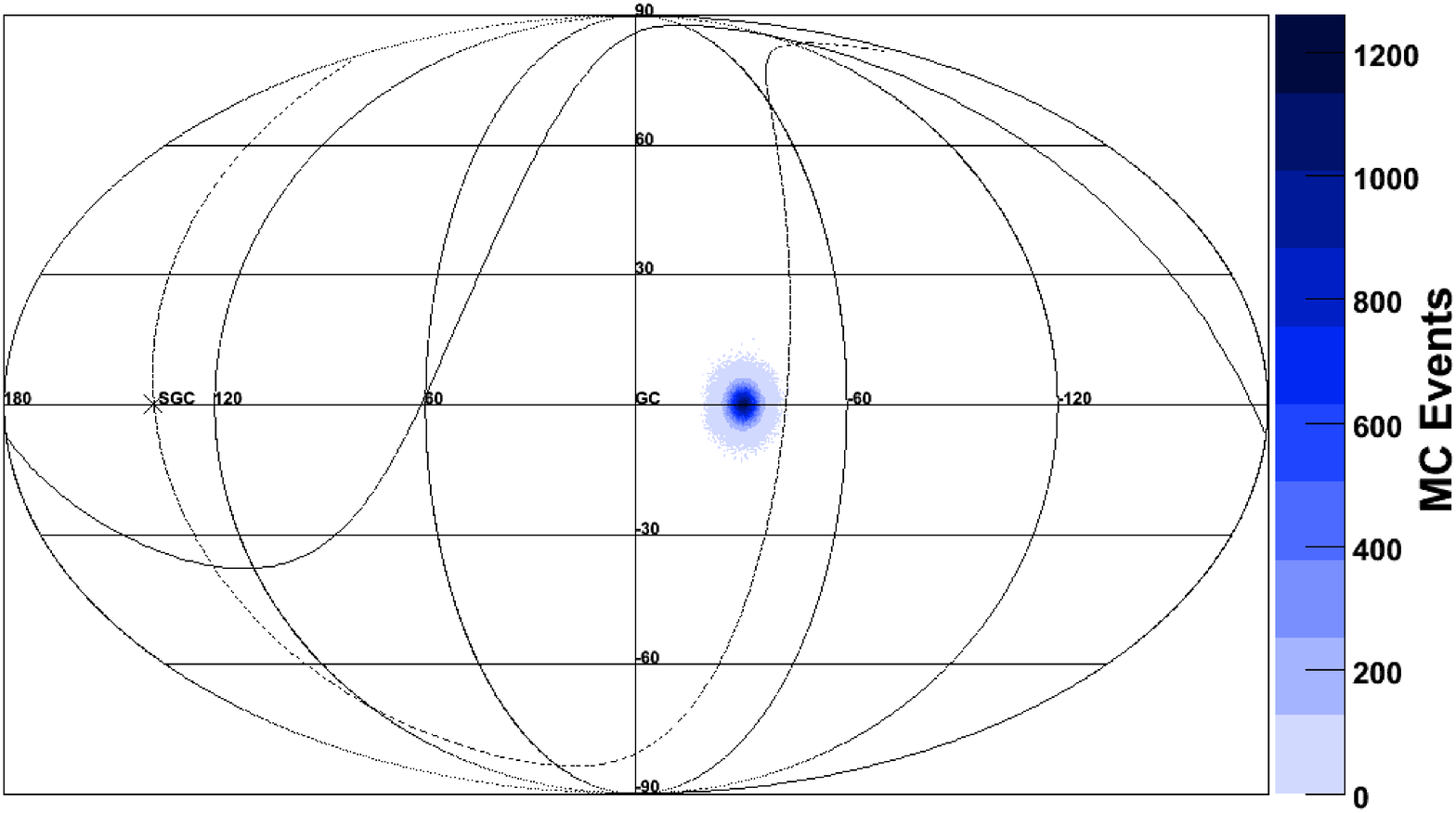} & 
    \includegraphics[width=0.50\linewidth]{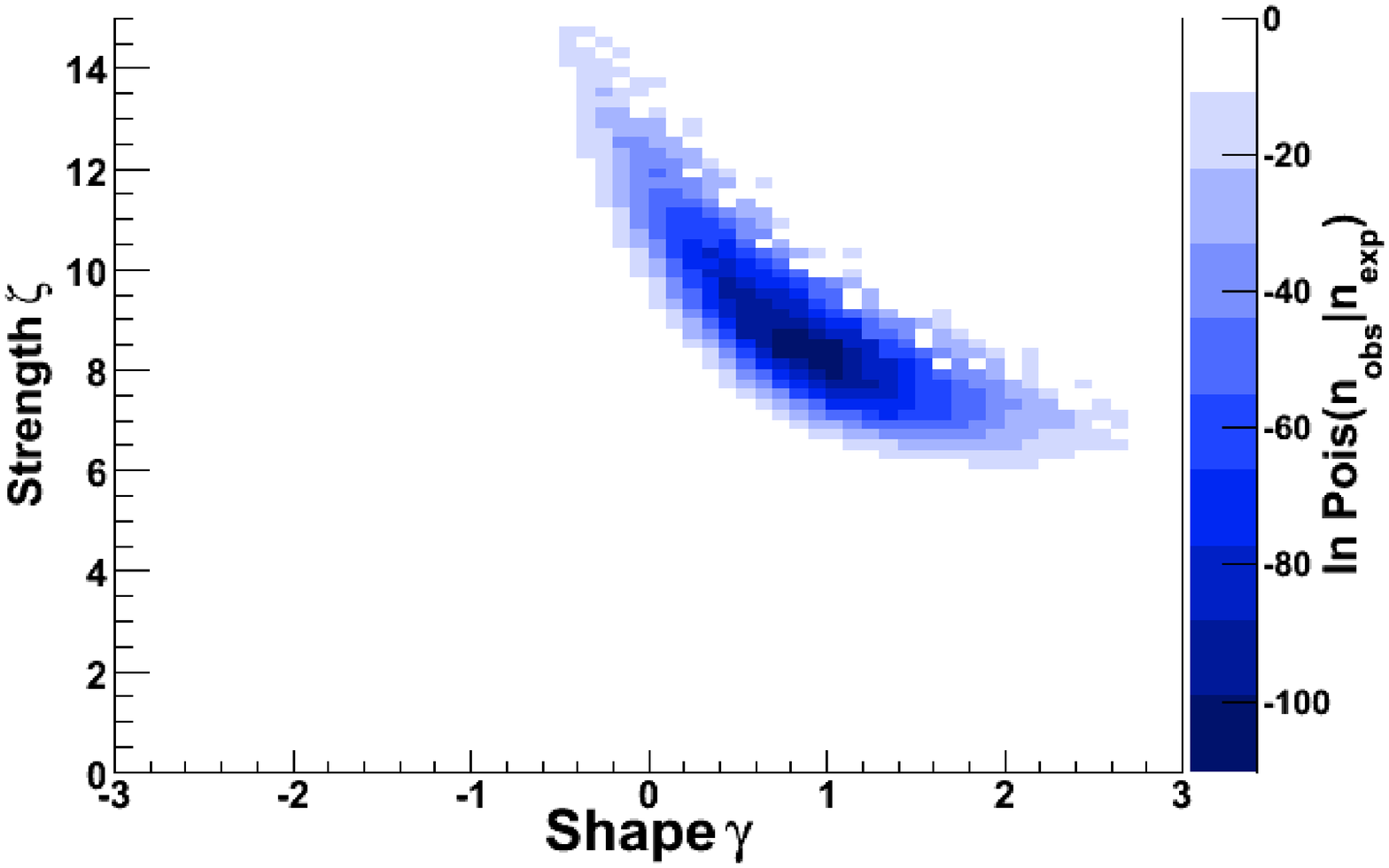} \\ 
    \includegraphics[width=0.50\linewidth]{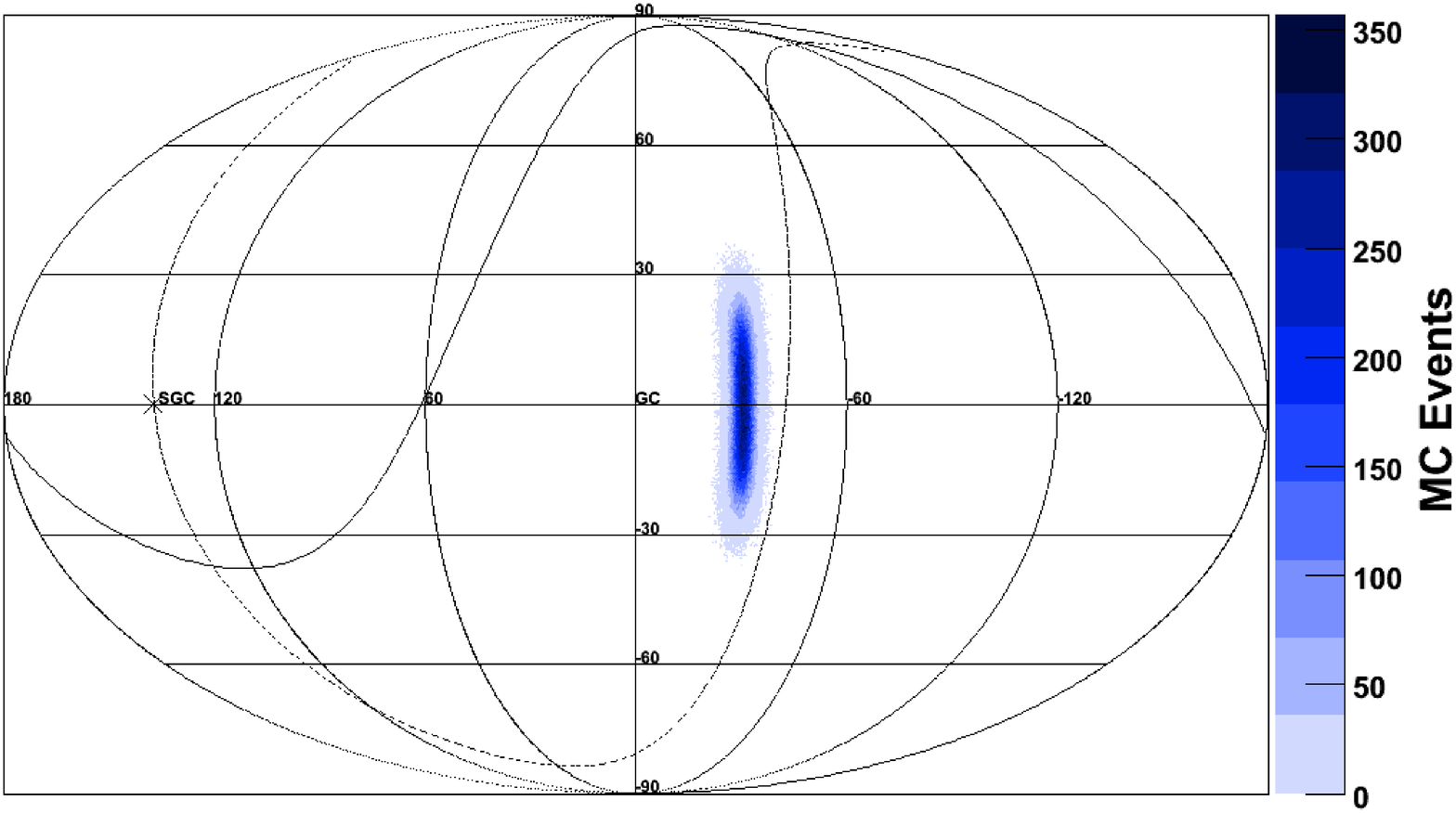} & 
    \includegraphics[width=0.50\linewidth]{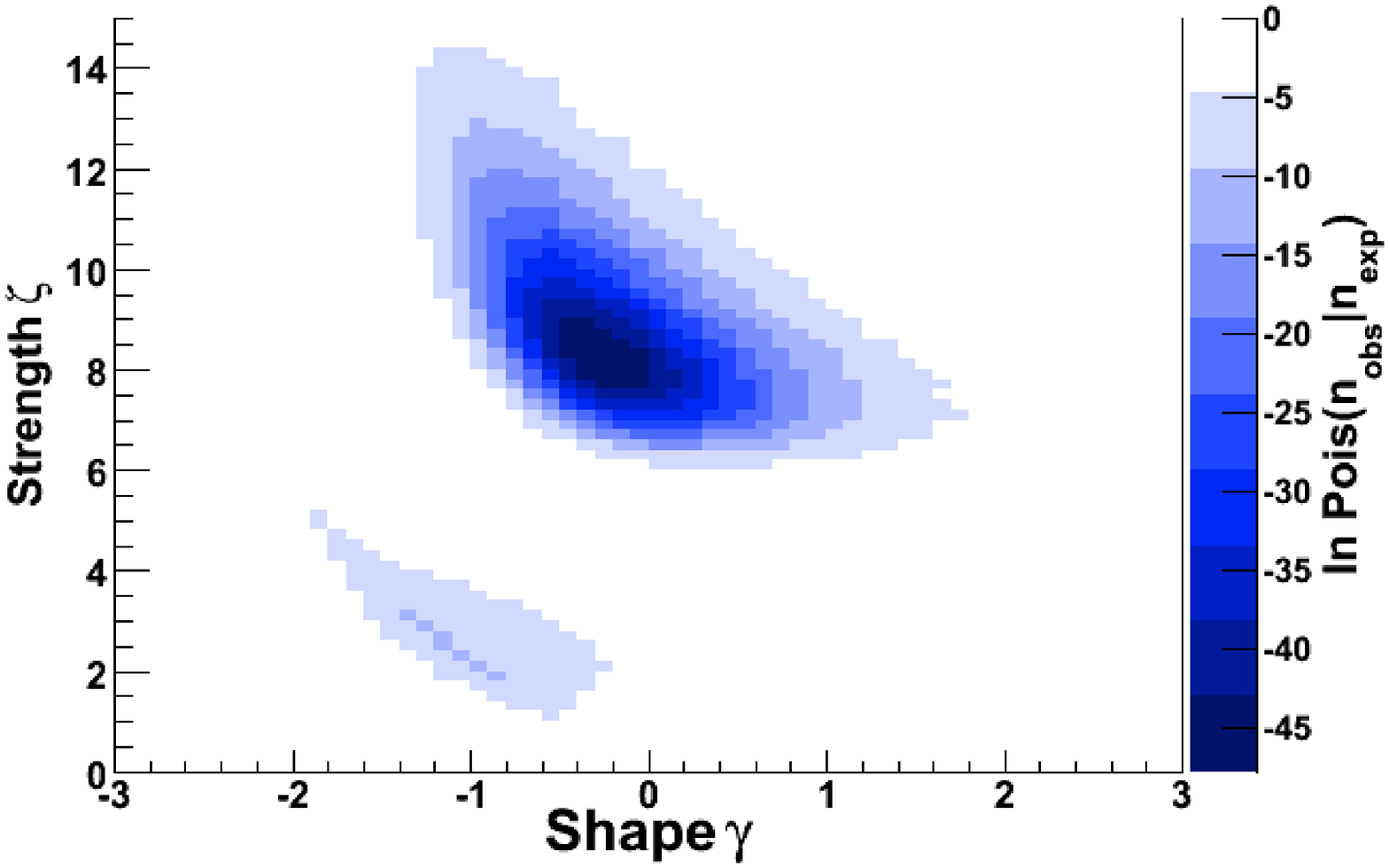} \\ 
  \end{tabular}
  \caption{\label{fig:points}
     {\it Left column:} Histogram of $10^{4}$ skies of 27 Monte-Carlo cosmic rays simulated from a single source 
     centered on $\{l,b\}=\{-30.0,0.0\}$ in galactic coordinates. 
     We use the Fisher-Bingham distribution\cite{Kent} on the sphere with $\kappa=400.0$ to generate these events.
     For the spherically symmetric point-like (top) distribution we use $\beta=0.0$.
     For the elliptically shaped (bottom) distribution we use $\beta=200.0$ with the major axis pointing perpendicular to the galactic plane. 
     See \cite{Kent} for a detailed description of the parameters $\kappa$ and $\beta$.
     {\it Right column:} The ensemble average (over all $10^4$ sets of 27 event skies) of the $\ln P(\nobs | \nexp)$ parameter space of the S-S method 
     for the point-like (top) and elliptically shaped (bottom) toy anisotropies. 
     In the bottom (right) panel one can see the relatively small deficit of triplets generated from the source with $\gamma \sim 1$ and $\zeta \sim 2$ 
     in addition to the large excess of triplets with $\gamma \sim 0$ and $\zeta \sim 8$. 
     The deficit arises from the non-uniform isotropic exposure of Auger\cite{Cronin:2007zz,Abraham:2007si} and the excess from the simulated source. 
     Both features contribute to the pseudo-likelihood where no distinction is made between excess and deficit of triplets. 
     These two signals can be consistently detected with both the 2-Pt and S-S methods. 
  }
\end{figure*}

It is convenient and statistically descriptive to work with a {\it shape}, $\gamma$, and a {\it strength}, $\zeta$, parameter\cite{refFisher}; 
\begin{linenomath*} 
  \begin{equation}
    \gamma = \lg \left\{ \frac{\lg(\tau_{1}/\tau_{2})}{\lg(\tau_{2}/\tau_{3})} \right\} \label{equ:shape} \\
  \end{equation}
\end{linenomath*}
\begin{linenomath*} 
  \begin{equation} 
    \zeta = \lg(\tau_{1}/\tau_{3}) \label{equ:strength}
  \end{equation}
\end{linenomath*}
As $\zeta$ increases from 0 to $\infty$ the events in the triplet become more concentrated. 
Generally, as $\gamma$ increases from $-\infty$ to $+\infty$ the shape of the triplet transforms from 
elliptical, i.e. strings, to symmetric about $\vec{u}_{1}$, i.e. point source. 
See the right panel of \figref{fig:trip} for a schematic representation.
In \figref{fig:points} we show the how the variation of the ellipticity of a source on the galactic sphere effects the shape-strength parameter space. 

To compute the test statistic $\SP$ using this method we sum over sixty bins for $\gamma \in [-3.0, 3.0)$ and 
seventy-five bins for $\zeta \in [0.0, 15.0)$, i.e. 
$\SP = \sum_{\gamma \zeta} \lnPss$.
We have checked that this parameter range is sufficient to cover event sets like those expected by Auger and that little is gained by 
enlarging the range. 

\section{Results} \label{sec:res}
In order to gain confidence in, and intuition about, the S-S method we apply it (and the 2-Pt correlation) 
to an astro-physically motivated simulated (mock) data set in \secref{ssec:ms}.
The results of applying the S-S and 2-Pt methods to the 27 most energetic events from Auger\cite{Cronin:2007zz,Abraham:2007si} 
are reported in \secref{ssec:ad}.

\subsection{Mock Signals} \label{ssec:ms}
In weighing the effectiveness of a method for rejecting the isotropy hypothesis $\ho$ for a given CR sky 
we are interested in the probabilities for two types of testing errors\cite{refpdg}.
A type I error is the probability $\tal$ (significance or $p$-value) of rejecting $\ho$ given that $\ho$ is {\it true}; 
in practice it should be chosen a priori. 
In this analysis we choose the 1\% significance level. One percent is arbitrary and is chosen to be the same 
as the value used in \cite{Cronin:2007zz,Abraham:2007si}. One could choose, for example, $0.1\%$ but this would require 
more data and/or a higher fraction of anisotropic signal to be detected.
For each method this choice corresponds to a unique $\SP^{\tal}$; 
we determine $\SP^{\tal}$ such that the ratio of the number of isotropic skies with $\SP$ less than
$\SP^{\tal}$ is $\tal=1\%$. We use the $\Nskies$ isotropic skies to determine the upper bound of the signal region of likelihood space.

A type II error is the probability, $\tbl$, of accepting $\ho$ (i.e. of rejecting the mock, or toy, signal hypothesis $\hs$) 
given that $\ho$ is {\it false}. 
This value is dependent on the choice of $\hs$. 
As we are interested in the effectiveness of accepting the signal hypothesis, we use the quantity $1-\beta$, called the {\it power} of the method\cite{refpdg}.
By applying the ensemble of each mock signal to each method we estimate the power as the ratio of mock signal skies 
with $\SP < \SP^{\tal}$ to the total number of mock skies. 
As a heuristic measure we will describe a method's power as ``good'' if it is at least 90\%, 
i.e. a high probability ($1-\tbl > 0.90$) to observe an anisotropy when there is an anisotropy in the data, 
and questionable if it is less than $90\%$.

\begin{figure*}[htp]
  \begin{tabular}{cc}
    \includegraphics[width=0.50\linewidth]{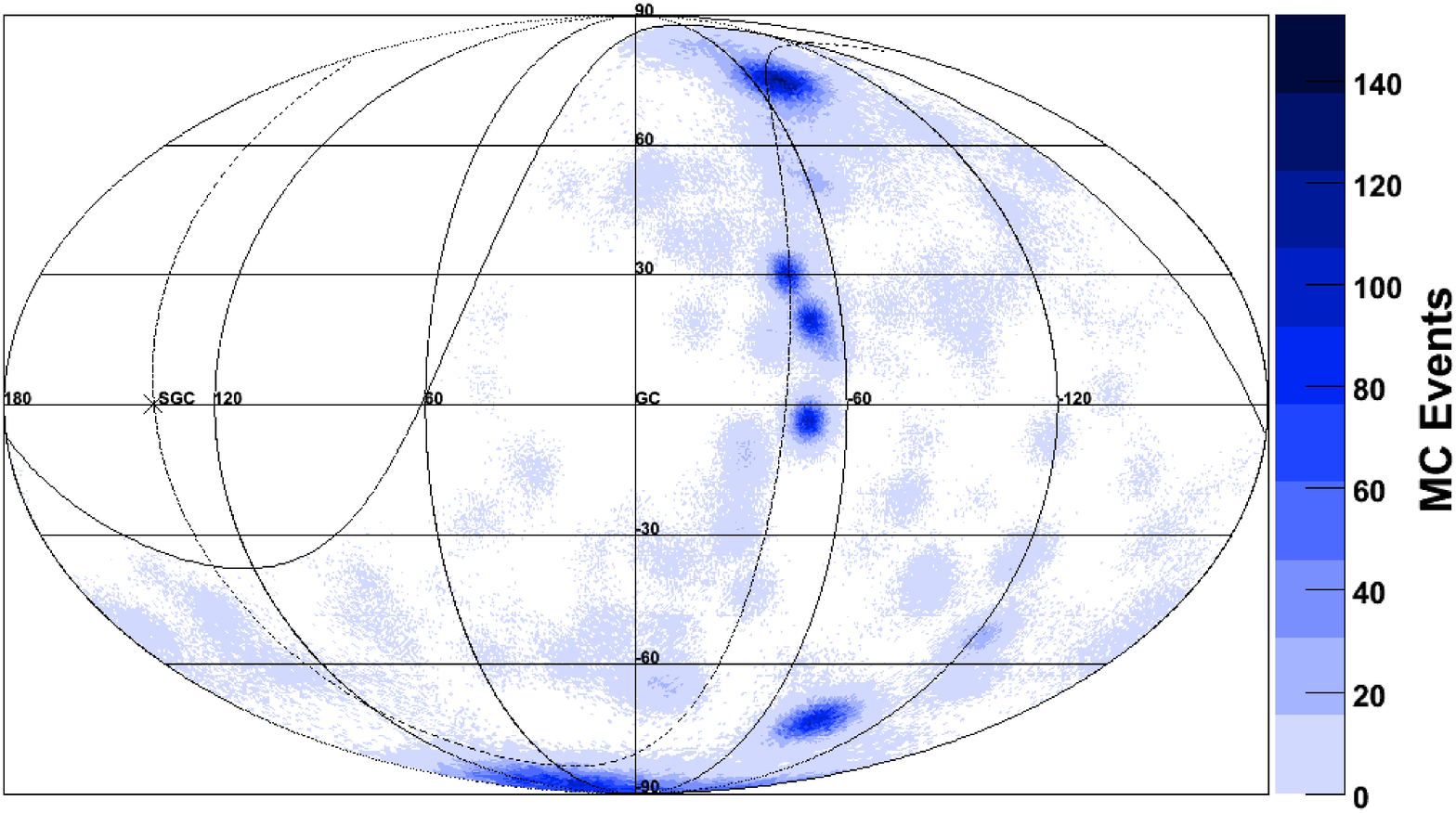} & 
    \includegraphics[width=0.50\linewidth]{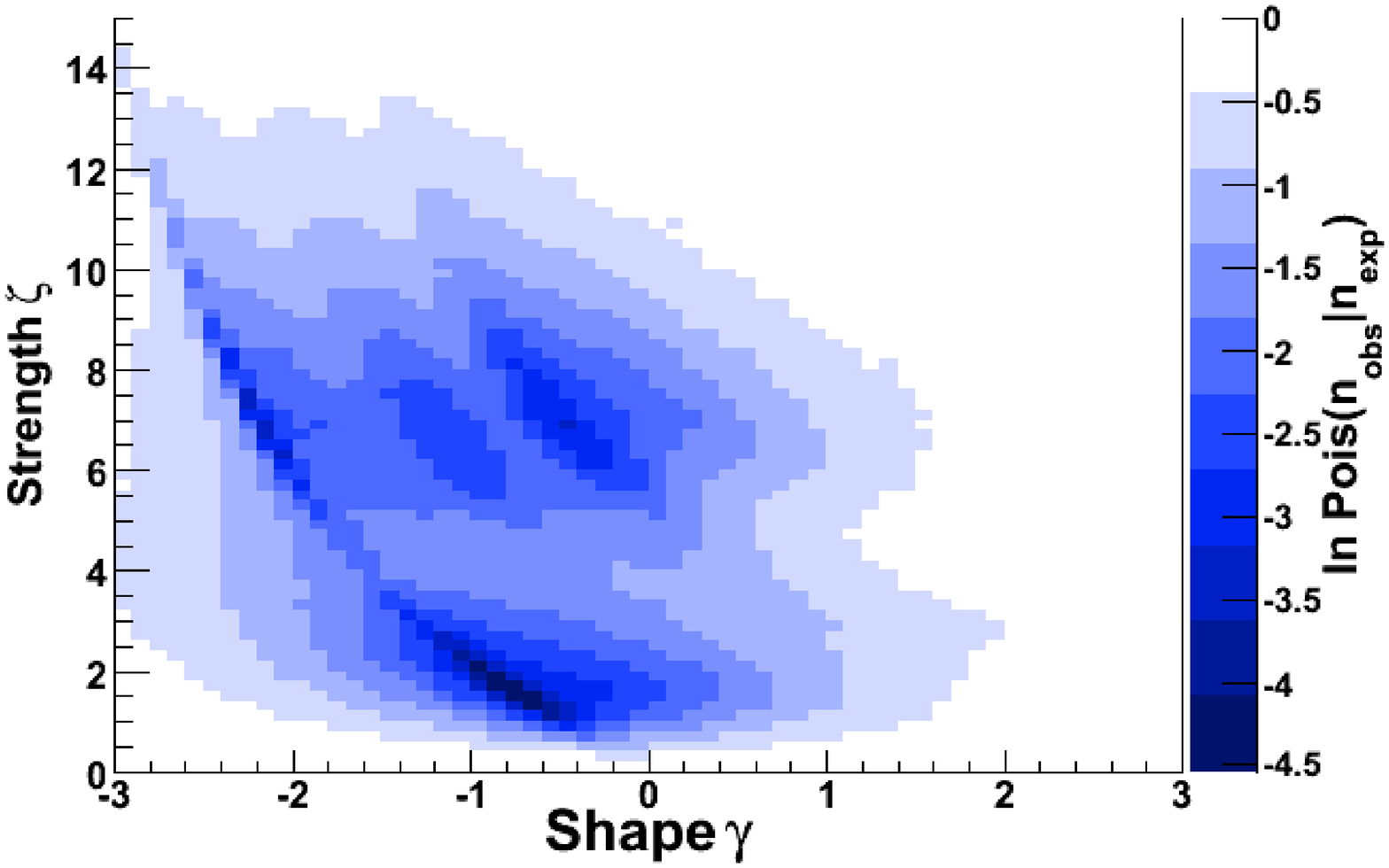} \\ 
    \includegraphics[width=0.50\linewidth]{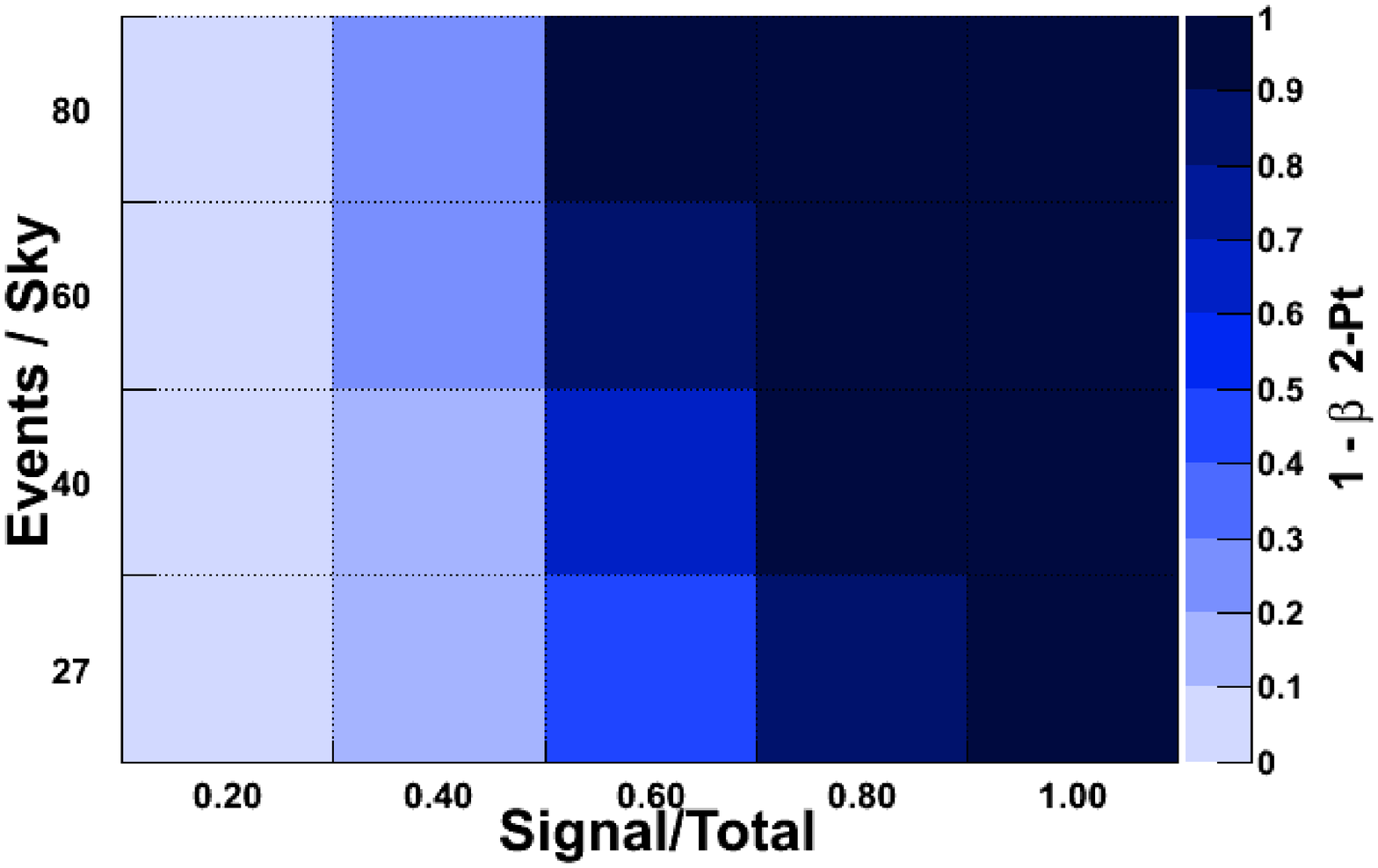} & 
    \includegraphics[width=0.50\linewidth]{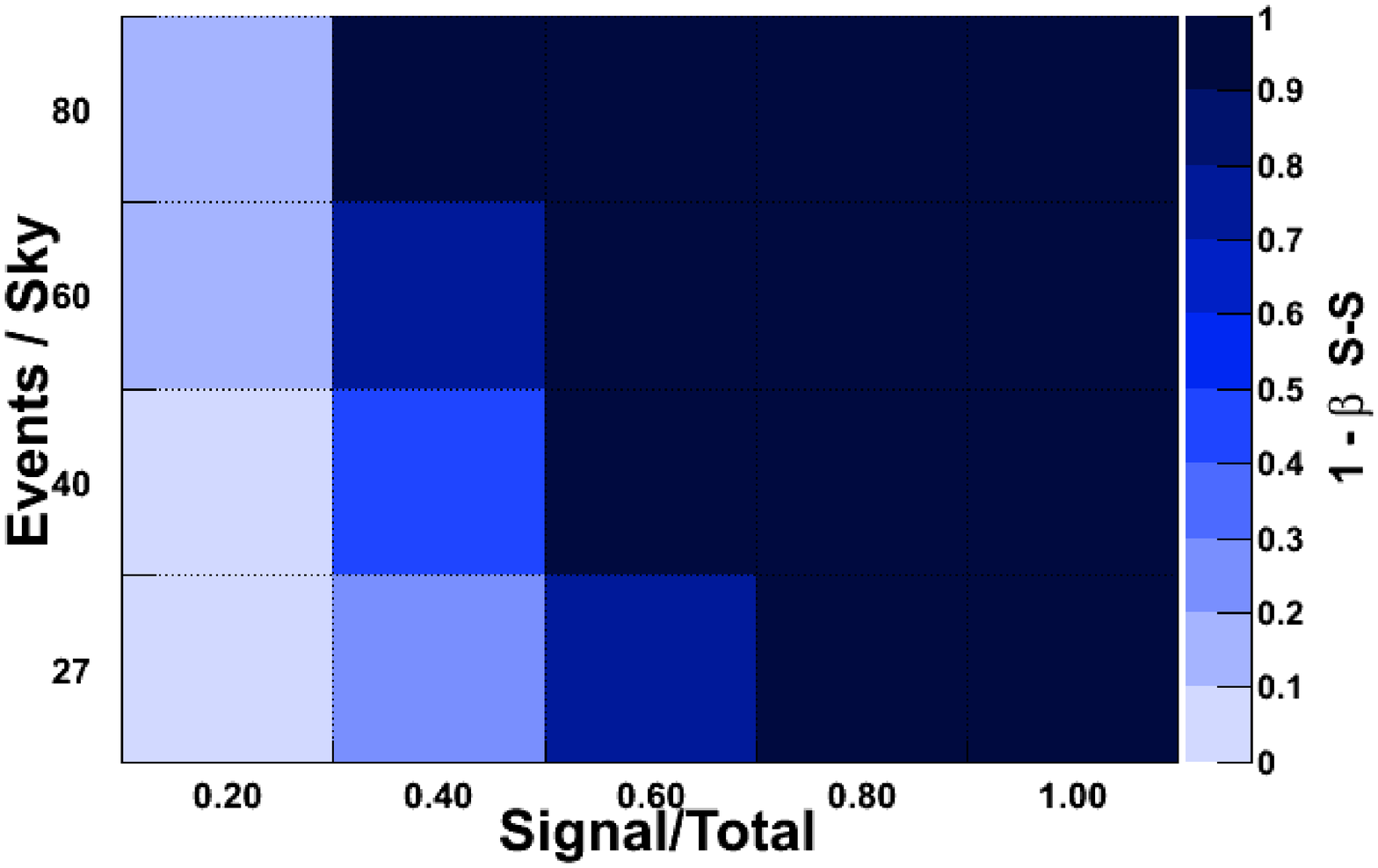} \\ 
  \end{tabular}
  \caption{\label{fig:skymaps}
     {\it Top left:} Histogram of $10^{4}$ skies of 27 Monte-Carlo cosmic rays simulated from the VCV\cite{refVCV} catalog.
     We select objects with redshift $z_{max} \leq 0.020$ and they are weighted by $1/z^2$. 
     Each simulated  CR is drawn from a collection of 2D-Gaussian probability distributions centered on the  
     the catalog sources, with deviation $\sigma=3\degree$.
     (See \secref{ssec:ms}.)
     {\it Top right:} Using the VCV catalog we plot the ensemble average of the $\ln P(\nobs | \nexp)$ parameter space of the S-S method. 
     {\it Bottom row:} Using the VCV ensemble files we can study the detection power $1-\beta$ as a function of the 
     number of events per sky and fraction of each sky containing signal events using both the 2-Pt ({\it left}) 
     and the S-S ({\it right}) methods.
  }
\end{figure*}

\begin{figure*}[htp]
  \begin{tabular}{cc}
    \includegraphics[width=0.50\linewidth]{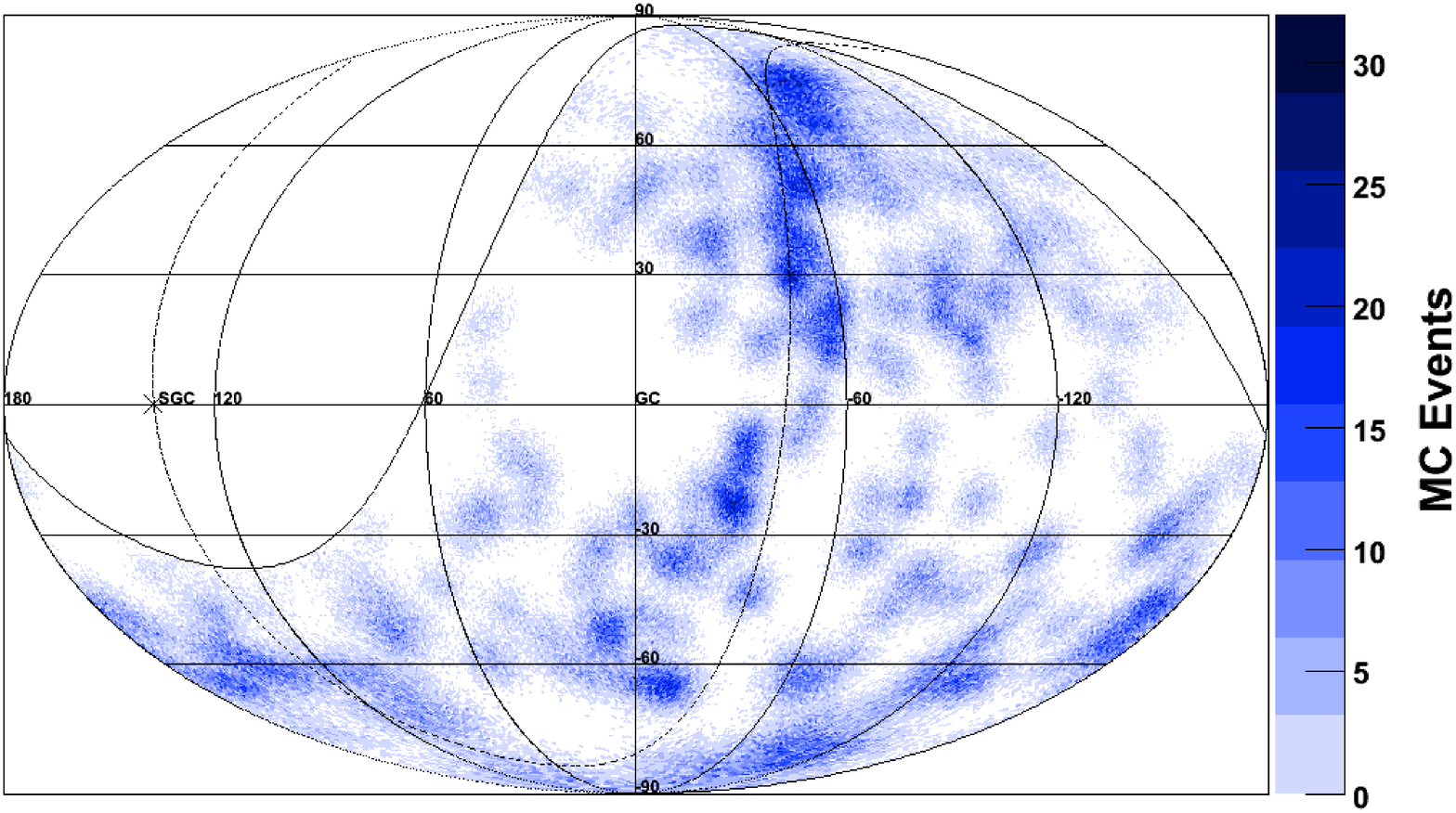} & 
    \includegraphics[width=0.50\linewidth]{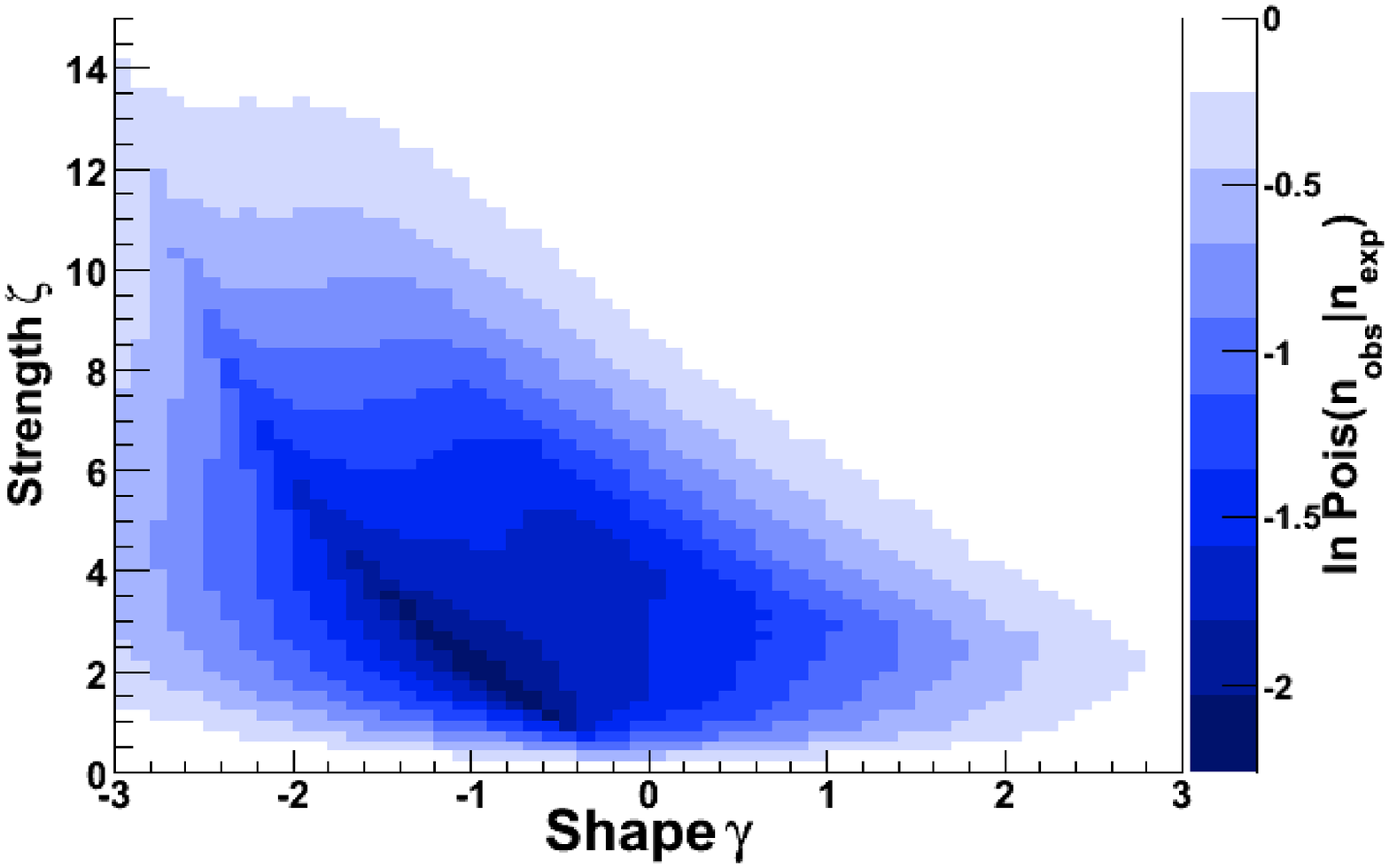} \\ 
    \includegraphics[width=0.50\linewidth]{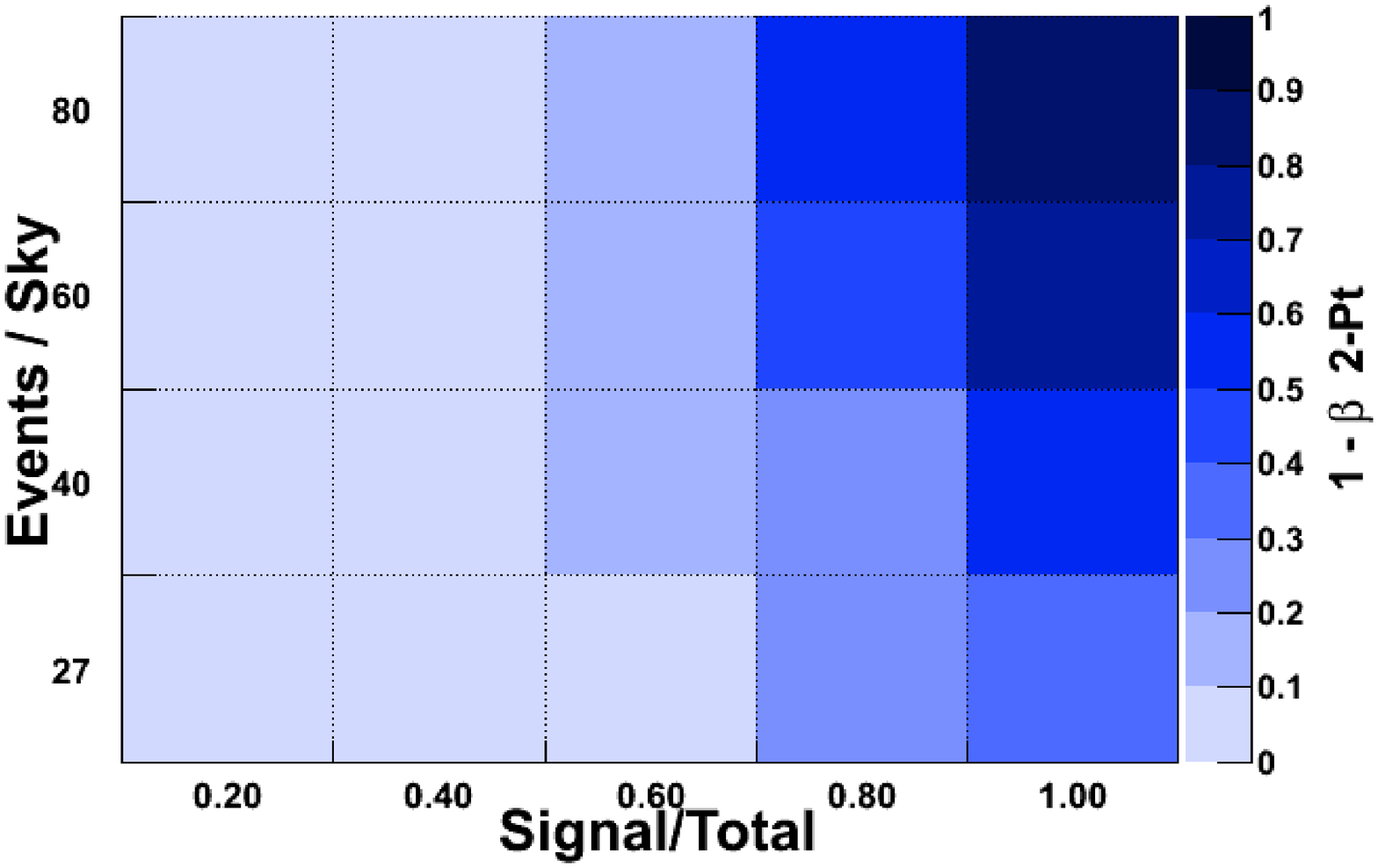} & 
    \includegraphics[width=0.50\linewidth]{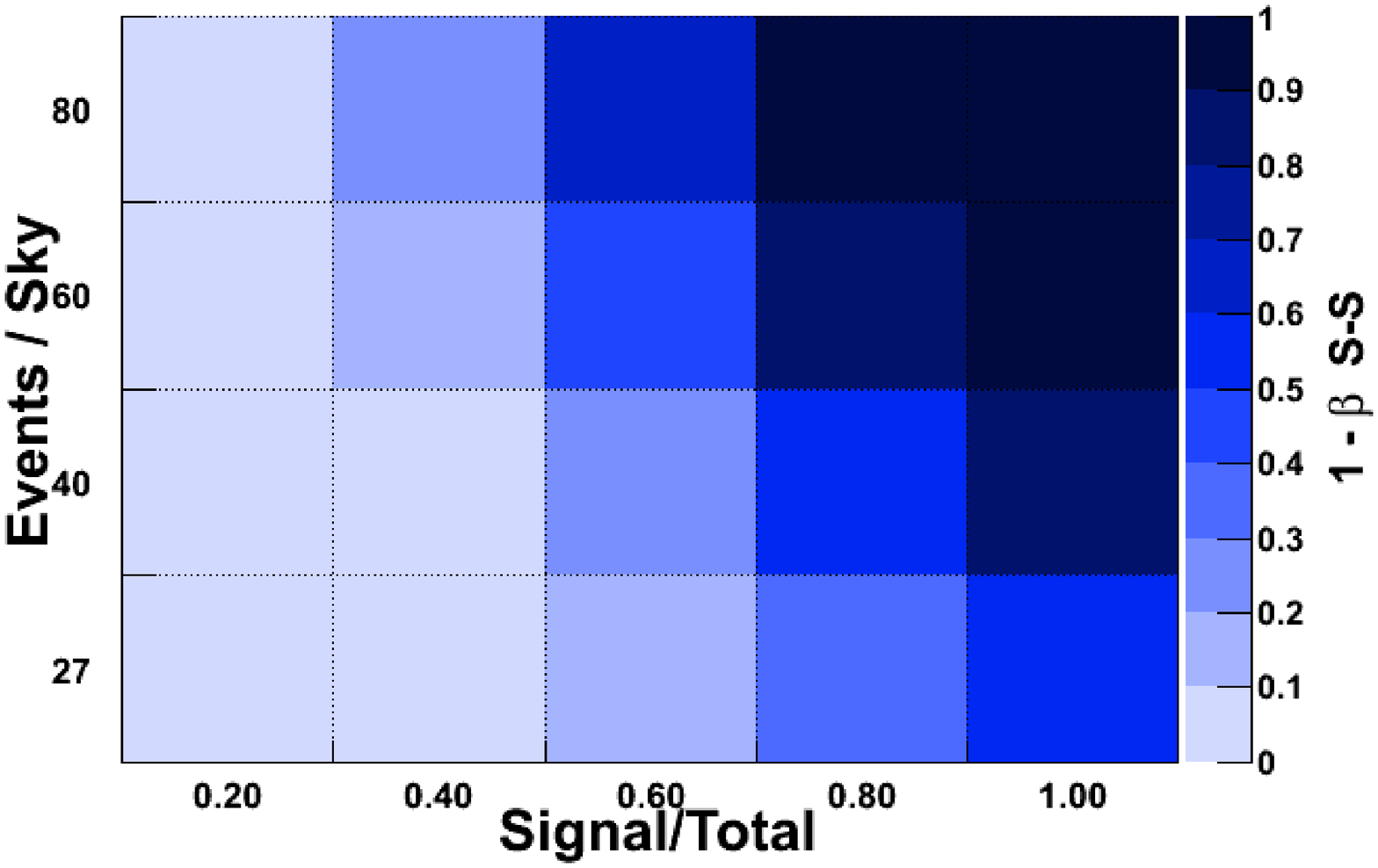} \\ 
  \end{tabular}
  \caption{\label{fig:skymapsflat}
     {\it Top left:} Histogram of $10^{4}$ skies of 27 Monte-Carlo cosmic rays simulated from the VCV\cite{refVCV} catalog.
     We select objects with redshift $z_{max} \leq 0.020$ and they are not weighted. 
     Each simulated  CR is drawn from a collection of 2D-Gaussian probability distributions centered on the  
     the catalog sources, with deviation $\sigma=3\degree$.
     (See \secref{ssec:ms}.)
     {\it Top right:} Using the VCV catalog we plot the ensemble average of the $\ln P(\nobs | \nexp)$ parameter space of the S-S method. 
     {\it Bottom row:} Using the VCV ensemble files we can study the detection power $1-\beta$ as a function of the 
     number of events per sky and fraction of each sky containing signal events using both the 2-Pt ({\it left}) 
     and the S-S ({\it right}) methods.
  }
\end{figure*}

Of significant physical interest is the ability of these catalog {\it independent} methods to detect signals 
generated from a catalog {\it dependent} map. 
To this end, we have studied simulated data sets generated from subsets of the 
VCV\cite{refVCV} galaxy catalog.
We consider only galaxies with reshift $z \leq  0.02$ and we weight each galaxy either by a $1/z^2$ acceptance factor or not at all. 
We simulate events arriving from these galaxies with a random component given by a two dimensional 
Gaussian centered on the galaxy and with deviation $\sigma=3\degree$. 
These choices are arbitrary in the sense that they describe some subset of nearby AGN with events 
smeared by a few times the angular resolution of Auger\cite{Abraham:2007si}. 
It should also be noted that the redshift weighted map (see \figref{fig:skymaps}) is highly anisotropic, 
consisting of a number of small to medium scale clumps on the celestial sphere, and is likely to yield multiple events per sky within these groups. 
In contrast, the unweighted VCV map (see \figref{fig:skymapsflat}) is notably more dispersed on the sphere. 

The true CR data is likely to contain a {\it mixture} of background events and signal events. 
To explicitly study this dilution effect has on the power we separately construct 
mock ensembles in which each sky has a certain ratio, $r$, of signal events to the total number of events, 
with $r = 0.2,0.4,\ldots,1.0$.
Notice that, because our methods use all the triplets or doublets in a given sky, 
the mixture $\SP$ distributions are not a simple sum of the signal and isotropic $\SP$ distributions.

Detection power is also strongly effected by the number of (high energy) events in a sky. 
The effect can be similar to those of signal dilution in that the power is decreased. 
We generate ensembles of $\Nskies$ skies with $27$, $40$, $60$, and $80$ events per sky from the VCV catalog.
Results for all combinations of source {\it purity} and number of (mock) data events
are reported in Figures \ref{fig:skymaps} and \ref{fig:skymapsflat}.  
The dark blue regions in lower plots of \figref{fig:skymaps} show that at least $40-80$ events with $(60-40)\%$ signal is required to achieve a 
high detection power, $1-\beta \sim 90\%$, for the redshift weighted VCV maps. 
The un-weighted VCV maps in \figref{fig:skymapsflat} require a nearly pure signal and 60 or more events to have high detection power. 

In general, where one method is good (power, $1-\beta \gtrsim 90\%$) so is another; the methods are correlated. 
However, the S-S method consistently performs better than the two point correlation for the types of signals discussed here. 

\subsection{Auger Data} \label{ssec:ad}
It is of interest to apply these techniques to experimental data.
The largest public ultra high energy data set is the 27 events that form the basis of the Auger result reporting evidence for anisotropy 
(at the 1\% significance level) in the highest energy CRs\cite{Cronin:2007zz,Abraham:2007si}.  
The p-values obtained are: $p \sim 3\%$ for the 2-point method\footnote{We note that the 2-point method used 
here differs from the auto-correlation analysis performed on the 27 events in \cite{Mollerach:2009tp}.} and $p \sim 0.2\%$ for the S-S method.
Thus, of these two methods only the S-S method would pass a requirement of $p < 1\%$ as evidence of anisotropy.
Note that these events are known to be anisotropic -- by the methods described in \cite{Cronin:2007zz,Abraham:2007si,Mollerach:2009tp} -- 
and therefore the $p$-values reported here reflect only on the {\it statistical methods} described in this paper.

\section{Conclusion} \label{sec:conc}
In this paper we have introduce a shape-strength method for testing isotropy on the unit sphere. 
We have shown that this method uses pattern-descriptive parameters and can consistently out-perform the two-point correlation method. 
By simulating artificial and astrophysically motivated signals of various sizes and purity we can gauge how this method might perform on real data. 
The S-S method out performs the two-point method for all of our parameter choices.

The S-S method was found to detect the redshift weighted VCV toy 
signal (having significant small scale anisotropies) with at least $\sim 50\%$ signal purity and about $60$ events in $>90\%$ of the Monte Carlo simulations. 
We also wish to emphasize from the analysis of the diluted mock signals that when the signal to all ratio $r \gtrsim 50\%$ 
we can expect that a redshift weighted ``VCV-like'' CR signal should be identified with power $\gtrsim 50\%$ by both methods for 
data sets with $\gtrsim 40$ events. 
The unweighted VCV toy signal (which is more diffuse on the sphere) is only reliably detected with greater than 80 events and 80\% signal purity. 

In agreement with qualitative expectations, this analysis demonstrates quantitatively how {\it both} signal purity and the total number of events 
dramatically effect the signal detection power. 
Furthermore, while sources types with significant small scale anisotropy can be detected with modest signal purity and total number of events, 
analysis of more subtle anisotropies suggest that either high purity signals or, more likely, much more data are needed for reliable identification.

\section{Acknowledgements} \label{sec:ackno}
We wish to thank members of the Auger collaboration for generous feedback and review of this paper. 
We also wish to thank Tim Thomas and the University of New Mexico Center for High Performance Computing\cite{refHpc}
for their generous donations of CPU processing power. 
This work is supported by DOE grant DE-FR02-04ER41300.

\newpage

\newpage

\newpage

\newpage
\section{References}
\bibliography{libros}

\begin{thebibliography}{10}
\bibitem{poistail}
The tail (large $\nexp$) of the true distribution in each bin falls somewhat
  less steeply than a Poisson tail. This approximation is made to save
  computational time and complexity, and results in slightly conservative
  $p$-values.
\end{thebibliography}
\bibliographystyle{unsrt}

\end{document}